\documentstyle[prl,aps,epsf]{revtex}

\begin{document}
\draft
\title{Stationary Properties of a Randomly Driven Ising Ferromagnet}
\author{Johannes Hausmann$^{\dagger}$ and P\'al Ruj\'an$^{\dagger,\$}$}
\address{Fachbereich 8 Physik$^{\dagger}$ and ICBM$^{\$}$, Postfach 2503,
         Carl von Ossietzky Universit\"at,
         D-26111 Oldenburg, Germany}
 
\date{\today}
\maketitle
\begin{abstract}
We consider the behavior of an Ising ferromagnet obeying the
Glauber dynamics under the influence of a fast switching, random
external field. Analytic results for the stationary state 
are presented in mean-field approximation, exhibiting a novel type of
first order phase transition related to dynamic freezing.
Monte Carlo simulations performed
on a quadratic lattice indicate that many features of the
mean field theory may survive the presence of 
fluctuations.
\end{abstract}
\bigskip

Many information processing systems,
natural or artificial, have a macroscopic number of connected elements
subject to external stimuli changing faster than the characteristic
thermal relaxation time. However,  attempts
at handling nonequilibrium stationary states of such
systems have been made only recently \cite{GC}.
As illustrated by the randomly driven Ising model presented 
in this Letter, stationary states far from equilibrium might be many times more
effective in dynamically storing information than thermal equilibrium
states.

The Ising  ferromagnet in a time-dependent sinusoidally oscillating field has received
recently a lot of attention, from both a theoretical and experimental
point of view. On the theoretical side, Rao, Krishnamurthy, and Pandit 
\cite{hyst-theo1} have
presented a large $N$-expansion of the cubic $O(N)$ model in three
dimensions and calculated the critical exponents related to the area of
the hysteresis loop. The underlying dynamic phase transition has been
then studied within both mean-field \cite{hyst-theo2} and Monte Carlo
simulations \cite{hyst-mc1,hyst-mc2,th-mc3,hyst-mc3}. The theory presented
in this paper is a generalization of these ideas for the case when
the external field is generated by chaotic dynamics and/or is a random variable.

Let a spin system $\vec \mu=(s_1,s_2,\dots,s_i,\dots,s_N)$, $s_i=\pm 1$, 
be in {\it local}
thermal equilibrium and denote by $P(\vec \mu, t)$ the
probability of finding the system in state $\vec \mu$ at time $t$.
The Master Equation used first by Glauber \cite{glauber} for
defining a stochastic dynamics for the Ising model reads then
\begin{equation}
        {dP(\vec \mu;t) \over dt}  
         =  \sum_i^N w(\vec \mu|\vec \mu_i) P(\vec \mu_i;t) - P(\vec \mu;t) \sum_i^N w(\vec \mu_i | \vec \mu)  
        \label{mastereq}
\end{equation}
where $\vec \mu_i =(s_1,s_2,\dots,-s_i,\dots,s_N)$. 
$w(\vec \nu | \vec \mu)$ denotes the transition rate 
from configuration $\vec \mu$ into state $\vec \nu$. Note that $P(\vec \mu;t)$ can be represented as a $2^N$ dimensional
vector. It is useful to expand  $P(\vec \mu,t)$ in the orthonormal basis formed by all
possible spin products, $P(\vec \mu;t)= {1\over 2^N} \sum_{\alpha}^{2^N} \pi_{\alpha} \prod_{i \in \alpha} s_i$,
where $\pi_{\alpha}=\langle \prod_{i \in\alpha} s_i {\rangle}_t$ \cite{glauber}.
In this `spin-correlation basis' the time-dependent distribution is given by $\vec \pi(t)$. 
The spectrum of the Liouville operator $\hat {\cal L}$ 
\begin{equation}
        {d \vec \pi  \over dt}  =  - \hat {\cal L}_{B(t)} \vec \pi(t) 
        \label{mastereqop}
\end{equation}
is invariant under such orthogonal transformations.
The slowest relaxation time corresponds to the decay of the order parameter (the
total magnetization) and is denoted by $\tau_{sys}$.
The typical phonon-spin interaction time is $\tau_{spin-flip} \ll \tau_{sys}$.
In what follows we assume that the external field $B$
is a random variable sampled identically and independently at discrete
times $t_n = n\tau_B, \ (n=0,1,\dots)$ from the symmetric
distribution:
\begin{equation}
        \rho(B(t_n)) = {1\over 2} \delta(B(t_n) - B_0) + {1\over 2} \delta(B(t_n) + B_0)
        \label{rhoH}
\end{equation}
As long as $\tau_B \gg \tau_{sys} $ 
the spin system relaxes  to global thermal equilibrium.
The situation is very different if $\tau_{sys} \gg \tau_B \gg \tau_{spin-flip}$. Now
the stationary state is determined by the distribution of the external field. 
By integrating Eq. (\ref{mastereq}) over the $(t_n,t_n+\tau_B)$ intervals, one 
obtains a `coarse grained' discrete Master Equation
\begin{equation}
        { \vec \pi(t_n +\tau_B) - \vec \pi(t_n) \over \tau_B }  = - \hat {\cal L}_{B(t_n)} \vec \pi(t_n)       
        \label{cgmastereq}
\end{equation}
which still describes correctly the long-time behavior of Eq. (\ref{mastereq}).
Eqs. (\ref{rhoH}-\ref{cgmastereq}) map our problem into an iterated function system (IFS) \cite{IRF}. 
 
The invariant probability density ${\cal P}_s [\vec \pi]$ induced by the dynamics (\ref{cgmastereq})
satisfies the Chapman-Kolmogorov equation:
\begin{equation} 
        \label{CK}
        {\cal P}_s (\vec \pi)  = \int d\vec \pi ' {\cal P}_s(\vec {\pi '}) \int dB \ \rho(B) \
                                \delta(\vec \pi - 
                                      {\rm e}^{- \hat {\cal L}_{B} \tau_B}\ \vec {\pi '}) = \left[ \delta(\vec \pi - 
                                      {\rm e}^{- \hat {\cal L}_{B} \tau_B} \vec {\pi '})    \right ] 
                                 \equiv  \hat {\cal K} {\cal P}_s (\vec \pi)
\end{equation}
where $[A]$ denotes the dynamic average $\int d\vec \pi ' {\cal P}_s(\vec {\pi '}) \int dB \ \rho(B) A $ and
$\hat {\cal K}$ the Frobenius-Perron operator.

As an example we consider the mean field Ising model. 
In this case all multispin correlations factorize in the thermodynamic limit and
the stationary distribution depends solely on the total 
magnetization. The energy is defined as usual,
\begin{equation}
        \label{mfenergy}
        E = - {J\over N} \sum_{i\neq j} s_i s_j - \mu_B B \sum_i s_i
\end{equation}
where $\mu_B$ is the Bohr-magneton. For the transition rate $w(\vec \mu_i|\vec \mu)$ we use 
the Glauber-form 
\begin{equation} 
        \label{trprob}
        w(\vec \mu_i|\vec \mu) = {1\over {2 \alpha}} [1 - s_i {\rm tanh}({ K\over N} \sum_{j\neq i} s_j + H)]
\end{equation}
where $\beta = 1/k_BT$, $K=\beta J$, $H=\beta \mu_B B$ and $\alpha$ sets the time constant.
Applying Eq. (\ref{cgmastereq}) one obtains after performing the thermodynamic limit $N\to \infty$:
\begin{equation}
        m(t+1)  = {\rm tanh}(Km(t) + H(t)) ,
	\label{mfmap} 
\end{equation}
where time is measured in units of $\tau_B$. 
The field distribution Eq. (\ref{rhoH}) leads to the one-dimensional map
\begin{equation}
        \label{mfbimap}
    m(t+1) = \left\{ \begin{array}{lcl} 
                      {{\rm tanh}(Km(t) + H_0)}\ & {\rm with\ prob.} &\ {1\over 2} \\
                     & & \\
                      {{\rm tanh}(Km(t) - H_0)}\ & {\rm with\ prob.} &\ {1\over 2} \\
                        \end{array} \right.
\end{equation}
Since in the stationary state, Eq. (\ref{CK}), $[m^k(t+1)]=[m^k(t)]$, 
using Eq. (\ref{mfmap}) and simple algebraic manipulations 
we obtain that the $k$-th moment of the stationary magnetization is given by
\begin{equation}
        \label{mfmoments}
        [m^k] = \left[ \left({v+h \over 1 + v h}\right)^k \right]\ \ \ \ \ k=1,2,...
\end{equation}
where $v={\rm tanh}(K m)$ and  $h={\rm tanh}(H)$. For example, when expanding 
up to fourth order in  $h_0={\rm tanh}(H_0)$  we get for the second moment
\begin{equation}
        [m^2]  \simeq  {h^2_0 \over  1 - K^2 (1-4h^2_0+3h^4_0) } 
        \label{mom2} 
\end{equation}
Higher moments can be calculated recursively in full analogy to methods introduced in \cite{RFIM2}.

\begin{figure}
  \begin{center}
        \leavevmode
        \epsfysize=4.2truecm \epsfbox{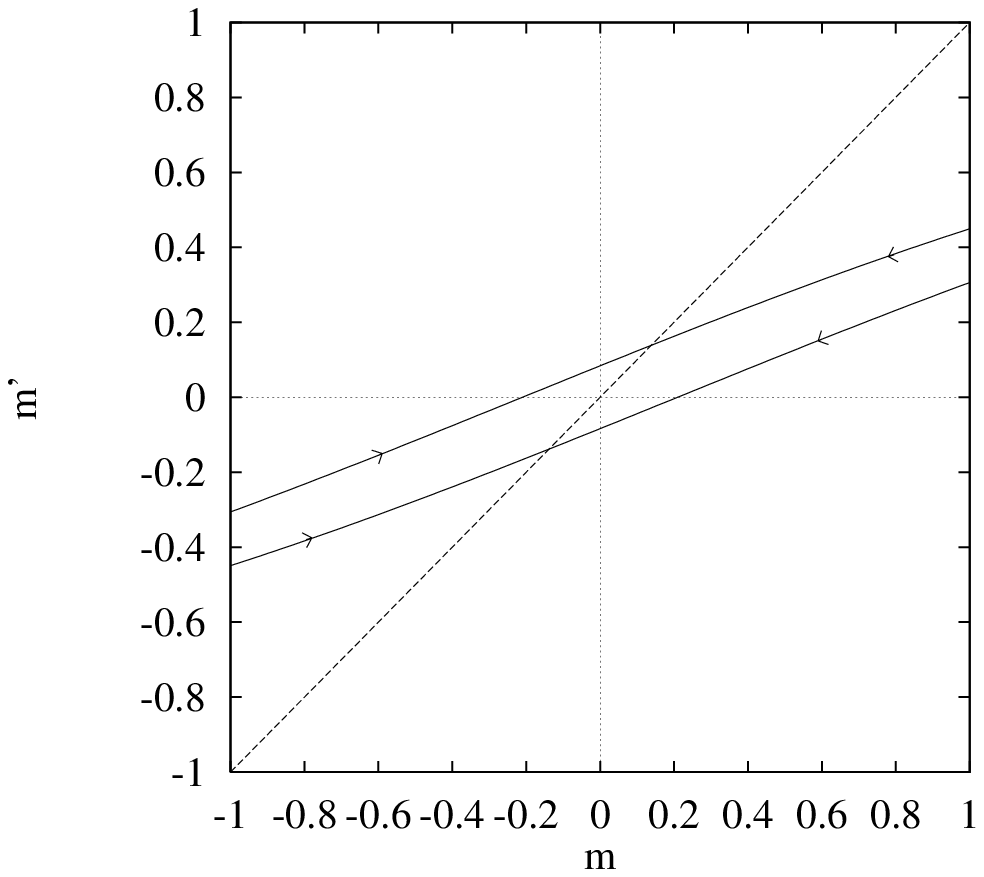}
        \epsfysize=4.2truecm \epsfbox{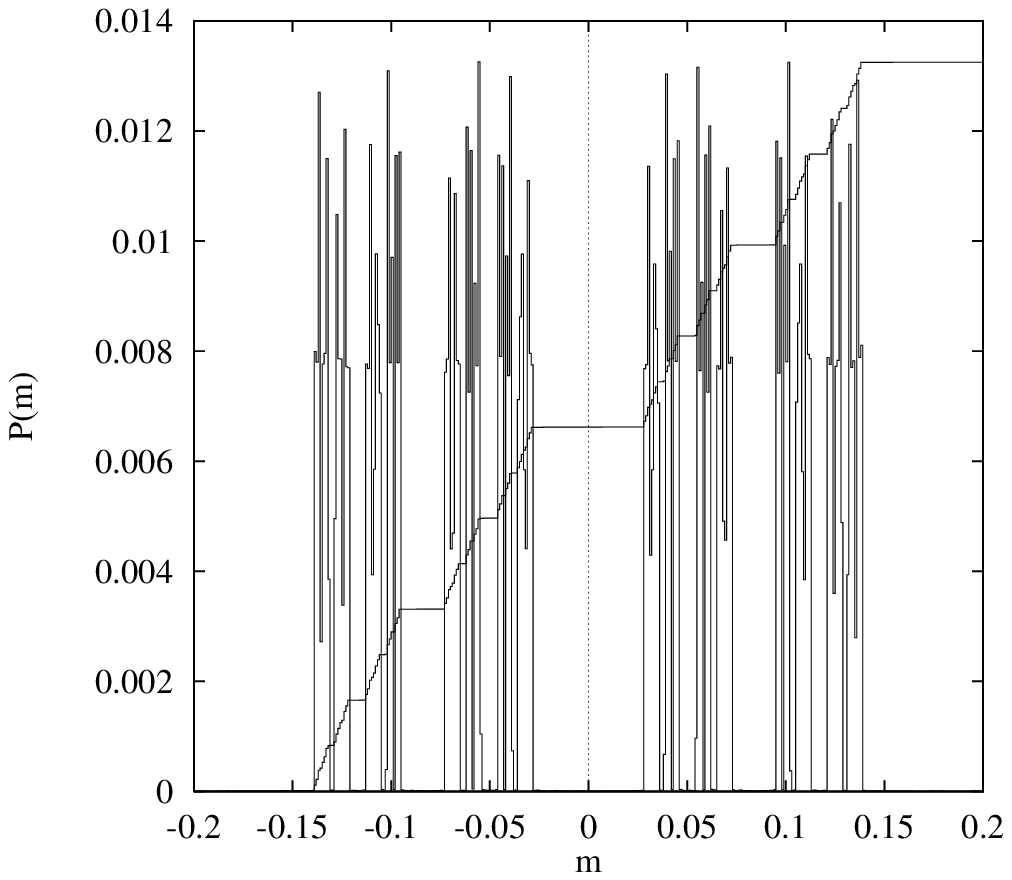}
  \end{center}
        \caption{\it Mean field map and the stationary 
        distribution in the paramagnetic phase ($K = 0.4$ and $H_0/K = 0.21$).
        }
        \label{snap0}
\end{figure}

\begin{figure}
  \begin{center}
        \leavevmode
        \epsfysize=4.2truecm \epsfbox{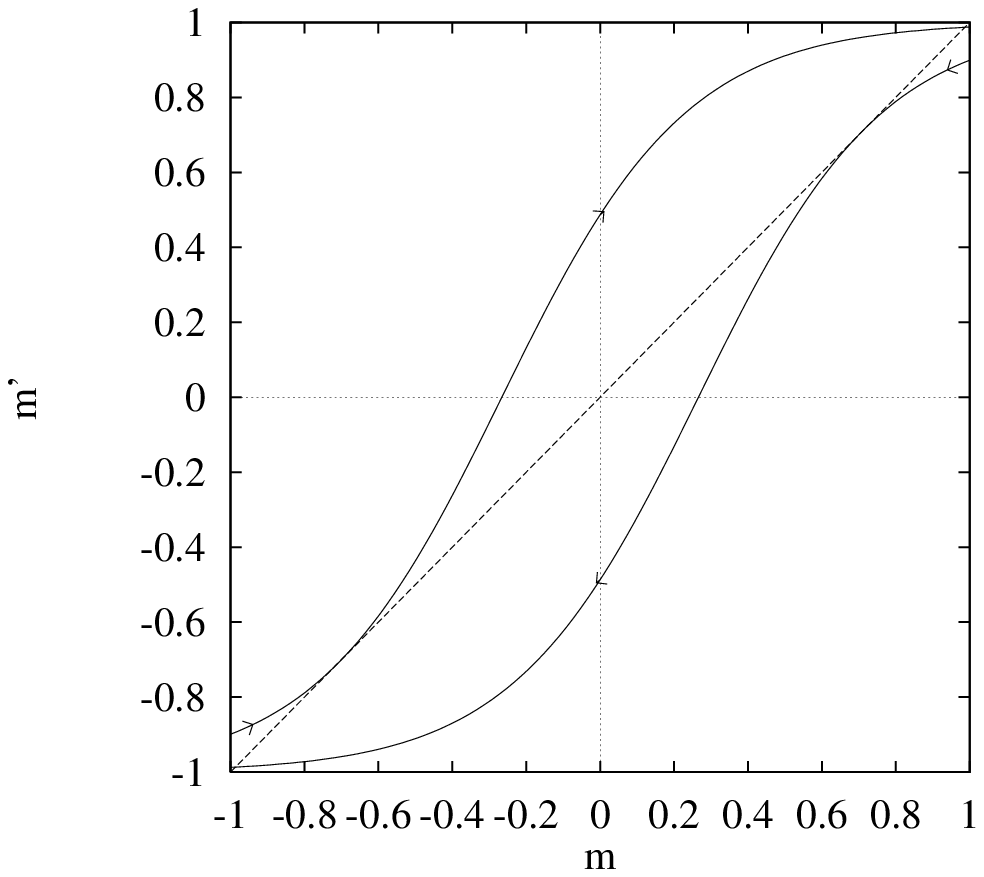}
        \epsfysize=4.2truecm \epsfbox{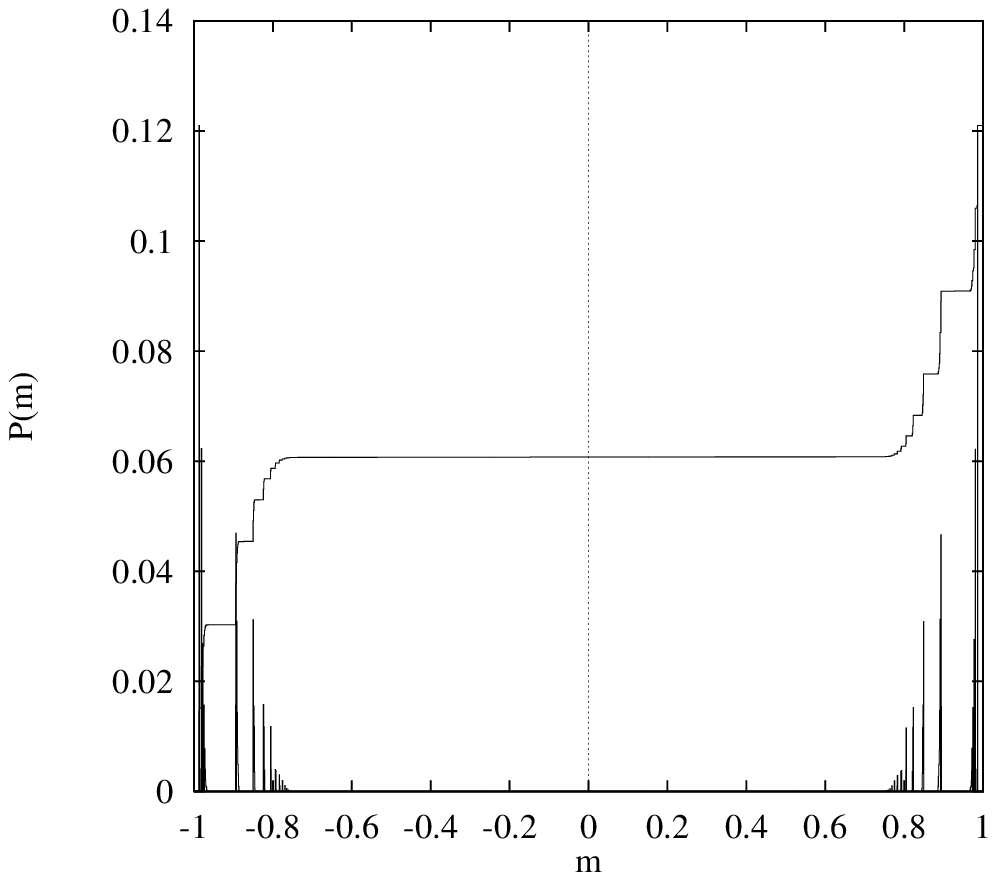}
  \end{center}
        \caption{\it Same as in Fig. 1 but close to the
            critical field value ($K = 2.0$ and $H_0/K = 0.266$).
            Two disjoint distributions are created around the stable fixed points,
            a repellor in the middle.}
        \label{snap2}
\end{figure}

\begin{figure}
  \begin{center}
	\leavevmode
        \epsfysize=5.0truecm \epsfbox{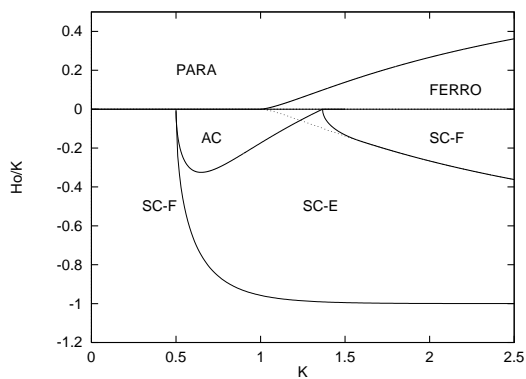}
  \end{center}
        \caption{\it  Mean field phase diagram. The upper part $(H_0>0)$ shows the
          border between the para- and ferromagnetic phase. In the lower part
          $(H_0<0)$ the regions
          denoted by SC-F, SC-E correspond to  
          a singular-continuous invariant density 
	  with fractal and Euclidean support, respectively,
          while in the AC-region the density is absolutely-continuous. 
          Note that the diagram is actually symmetric in $H_0$.}
        \label{mfphgddiag}
\end{figure}

The two graphs below, Figs. \ref{snap0} and \ref{snap2}, show how the map 
Eq. (\ref{mfbimap}) changes between 
high and low temperature. For further use let us denote by 
$m_1,\ m_2$, and $m_3$ the possible
fixed points of the equation $m=\tanh(Km+H_0)$ in descending order.
The stationary magnetization distribution 
undergoes a tangential bifurcation at the critical field
\begin{equation}
        \label{pht-first}
        H_c = {1\over 2 } {\rm ln}{1-m^{\dag}\over 1+m^{\dag}} + Km^{\dag}
\end{equation}
where $m_2 = m_3 = m^{\dag} = \pm \sqrt{K-1\over K}$ for $K > 1$. The 
corresponding phase diagram
is shown in the upper part of Fig. \ref{mfphgddiag}. Below the phase transition 
the stationary magnetization
distribution bifurcates into two symmetric, stable 
`spontaneous magnetization distributions'
and a central repellor. The phase transition is first order, the average 
stationary magnetization jumps at the phase border.

A different kind of transitions are related to
the analytic structure of the invariant density.
Following the notation introduced in \cite{radons}, one can identify
a singular-continuous density with fractal support (SC-F) in both the
paramagnetic and the ferromagnetic phase. 
When a gap opens between the upper and the lower branch of the map the 
invariant distribution has a fractal support with the capacity
dimension $d_0 < 1$. The border of the (SC-F) region is given by $Km_1 = H_0$ in the
para- and $K(m_1+m_3)=2H_0$ in the ferromagnetic phase.  In the
region between $d_0 = 1$ and $d_{\infty}=1_-$ the distribution is
singular-continuous with Euclidean support (SC-E) \cite{radons}. Using the ideas developed in
\cite{evang}, we obtain $d_{\infty}=1$ if $K(1-m_1^2)={1\over2}$.
The density distribution is absolutely continuous (AC) 
if all generalized dimensions \cite{hp} equal one, $d_q=1$, $(q=0,\dots,\infty)$. 
These results are graphically summarized in the lower part of Fig. \ref{mfphgddiag}. 
 
The generalized free energy of such a driven system can be defined as $-\beta {\cal F} = \Lambda$, 
where $\Lambda$ is the largest Lyapunov exponent of the map Eq. (\ref{mfbimap}).
As expected, in the thermodynamic limit we obtain a 
dynamic average over the thermal mean field free energy at
magnetization $m$. After performing the average over $\rho(B)$ the 
generalized free energy per spin is given by
\begin{equation}
        \label{mffe}
        -{ \beta  {\cal F} \over N} = \int dm {\cal P}_s(m) {1\over 2} \ln 2 (\cosh(2 K m) + \cosh (2 H_0))
\end{equation}
Strictly speaking, Eq. (\ref{mffe}) is the average free energy. When considering a finite system or
a long but finite dynamic trajectory, the free energy is normally distributed. As shown in \cite{peter2} 
for the one dimensional random field Ising model,
in the SC-F region the multifractal spectrum can be directly related to the second cumulant of the 
free energy distribution. The arguments presented in \cite{peter2} apply also to our case, 
a broad multifractal distribution leads to large free energy fluctuations.

Consider now Fig. \ref{snap2} at negative
magnetization values. Close to but above the critical field (\ref{pht-first})
the upper branch of the map comes close but does not touch yet the diagonal. Let $n$ be 
the average number of iterations needed to pass through this region starting from
the lower corner by following the upper branch of the map. 
According to the general theory of type-I intermittency in one-dimensional maps,
$n \sim (H_0-H_c)^{-1/2}$. Since in each iteration step the probability of
jumping back along the lower branch is 1/2, the total time spent in the
lower corner is proportional to $2^n$, or
\begin{equation}
        \label{relexp}
        \tau \sim  2^n \sim 2^{c [H_0-H_c]^{-{1\over 2}} }
\end{equation}
with  $c \sim O(1)$ a constant. 
This behavior suggests a {\it dynamic freezing} transition (for more details see \cite{hr}).

\begin{figure}
  \begin{center}
    \leavevmode
    \epsfysize=4.2truecm \epsfbox{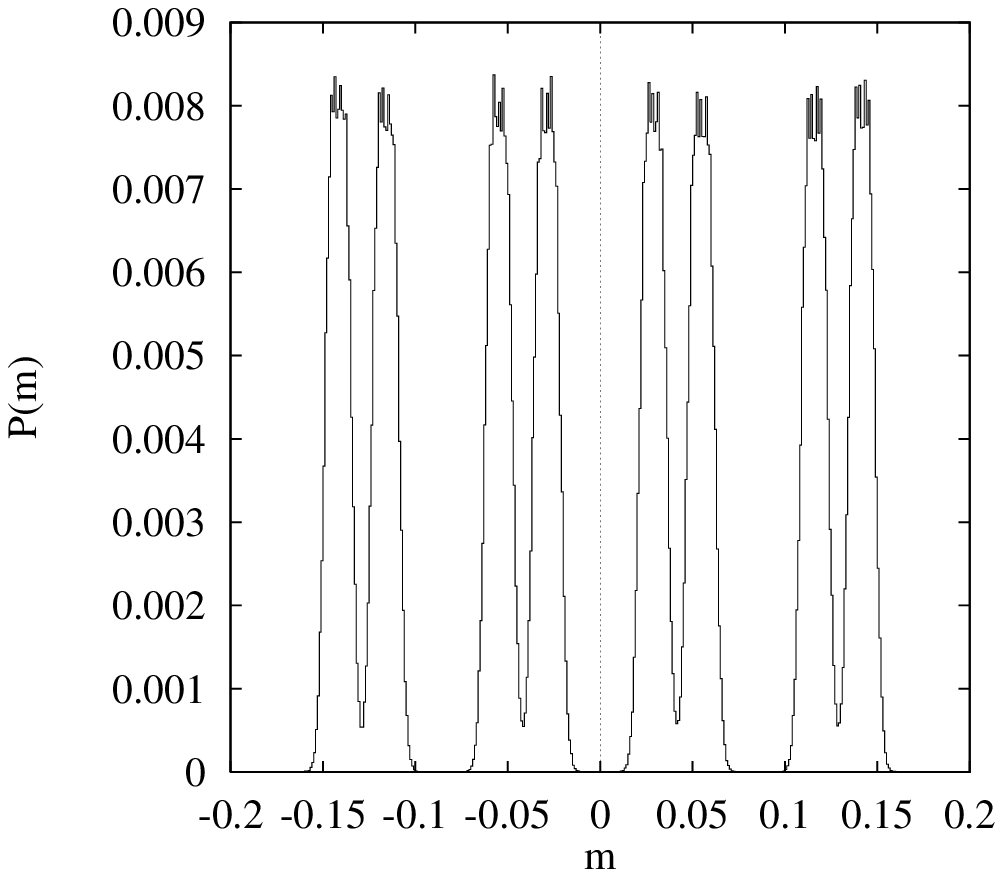}
    \epsfysize=4.2truecm \epsfbox{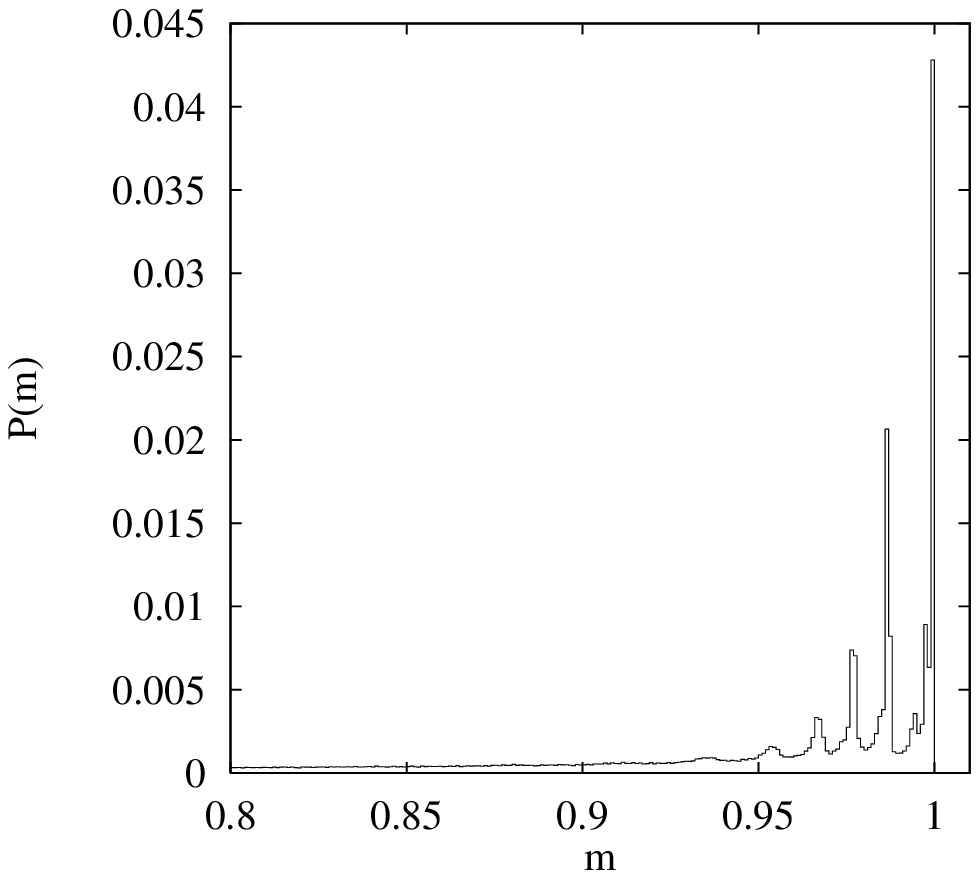}
  \end{center}
  \caption{\it Left: Magnetization distribution for the square lattice 
     RDIM averaged over eight different initial conditions.  $K/K_c=0.4,\ K_c=0.4407,\ H_0/K = 0.5$,   
    lattice size $415\times 415$. The simulation was run for $2 \cdot 10^{5}$ MCS. 
    Compare to Fig. 1. Right: Same parameters but $K=2K_c, \ H_0/K = 1$.
    Compare to Fig. 2.
    }
  \label{2d-mdist}
\end{figure}

In order to test the predictions of the mean field theory in a more realistic
setting, we performed Monte
Carlo simulations for an Ising model with nearest neighbor interactions
on a square lattice. 
The driving field is sampled from the distribution Eq. (\ref{rhoH}) after
each Monte Carlo step (MCS).
The left side of  Fig. \ref{2d-mdist} shows
the measured magnetization distribution in the paramagnetic phase, which is similar
to Fig. \ref{snap0}.
Below the critical temperature one obtains distributions similar to the ones
displayed on the right side of Fig. \ref{2d-mdist}, to be compared with Fig. \ref{snap2}.
Note that the critical field $H_c(K)$ is not a universal quantity and 
the square lattice values are different from the mean field ones. More details will
be published elsewhere.
Thermal fluctuations and finite
size effects wash out the fine structure of the multifractal
magnetization distribution predicted by the mean-field
theory. However, the sharp peaks and the presence of gaps indicate
that at least the main features of the magnetization distribution
are preserved in two-dimensions. 

Besides the theoretical interest in describing such systems, 
we believe that our predictions can be tested with recently developed experimental
techniques.
Dynamic magnetization measurements have been recently performed
in ultrathin Au(111)/Cu(0001)/Au(111) sandwiches 
or epitaxial Co/Au(111) films \cite{exp1,exp2,exp3}. Similarly,
hysteresis measurements on the ultrathin film
Co/Au(001) \cite{exp4} indicate that
below $T_c$ these systems undergo a dynamic phase transition
belonging to the Ising-universality class. 
More relevant to our theory, the time evolution 
of magnetization clusters can be optically recorded. The typical
relaxation times range from minutes to a few seconds with increasing
field amplitudes \cite{exp3}. 
This relatively slow relaxation rate allows for a simple 
experimental realization of the randomly driven external field. 

Ultrathin films are potential candidates
for magneto-optical storage devices. At well chosen control parameters the 
stationary magnetization distribution of the RDIM 
displays several well separated peaks. Hence, by coding appropriately
the time-sequence of field switches,
 one can - {\it in principle} - store locally 
more than two binary states. 

We are grateful to U. Ramacher and the ZFE, Siemens AG
for the SYNAPSE1/N110 neurocomputer, 
on which the Monte Carlo simulations were performed. 
This work was partly supported by the DFG through SFB 517.



\end{document}